\documentclass[conference, top=0.7in]{IEEEtran}
\usepackage[T1]{fontenc}
\usepackage{cite}
\usepackage[cmex10]{amsmath}
\usepackage{amsthm}
\usepackage{amssymb}
\usepackage{mathtools}
\usepackage[hidelinks]{hyperref} 
\usepackage{algorithmic}
\usepackage{textcomp}
\usepackage[usenames, dvipsnames]{color}
\usepackage{algorithm,algorithmic}
\usepackage{subcaption}
\usepackage{graphicx}  
\usepackage{url}
\usepackage{verbatim}
\usepackage[a4paper, top=0.75in,left=0.625in,right=0.625in,bottom=1in]{geometry}

\makeatletter
\newcommand{\ALOOP}[1]{\ALC@it\algorithmicloop\ #1%
	\begin{ALC@loop}}
	\newcommand{\ENDALOOP}{\end{ALC@loop}\ALC@it\algorithmicendloop}

\makeatother



\usepackage {tikz}
\tikzset{>=stealth}
\usepackage{xcolor,colortbl}

\newcommand{\mbf}[1]{\mathbf{#1}}
\newcommand{\mbb}[1]{\mathbb{#1}}

\newcommand{\mtt}[1]{\mathtt{#1}}

\newcommand{\RIS}{\mtt{R}}

\newcommand{\diag}[1]{\operatorname{diag}{#1}}

\DeclareMathOperator*{\argmax}{arg\;max}

\ifCLASSINFOpdf
\else
\fi
\usepackage{amsmath}
\begin{document}
%
\title{A Deep Reinforcement Learning Approach for Autonomous Reconfigurable Intelligent Surfaces}
%
%
%

\author{
    \IEEEauthorblockN{
        Hyuckjin~Choi\IEEEauthorrefmark{1},
        Ly~V.~Nguyen\IEEEauthorrefmark{2},
        Junil~Choi\IEEEauthorrefmark{1}, and
        A.~Lee~Swindlehurst\IEEEauthorrefmark{2}
    }
    \IEEEauthorblockA{
        \IEEEauthorrefmark{1}School of Electrical Engineering,
        Korea Advanced Institute of Science and Technology, Daejeon, South Korea
    }
    \IEEEauthorblockA{
        \IEEEauthorrefmark{2}Center for Pervasive Communications and Computing,
        University of California, Irvine, California, USA
    }
    
    \thanks{This work was supported in part by the National Science Foundation under grants CNS-2107182 and ECCS-2030029.}
}

\maketitle

\begin{abstract}
A reconfigurable intelligent surface (RIS) is a prospective wireless technology that enhances wireless channel quality. An RIS is often equipped with passive array of elements and provides cost and power-efficient solutions for coverage extension of wireless communication systems. Without any radio frequency (RF) chains or computing resources, however, the RIS requires control information to be sent to it from an external unit, e.g., a base station (BS). The control information can be delivered by wired or wireless channels, and the BS must be aware of the RIS and the RIS-related channel conditions in order to effectively configure its behavior. Recent works have introduced hybrid RIS structures possessing a few active elements that can sense and digitally process received data. Here, we propose the operation of an entirely autonomous RIS that operates without a control link between the RIS and BS. Using a few sensing elements, the autonomous RIS employs a deep Q network (DQN) based on reinforcement learning in order to enhance the sum rate of the network. Our results illustrate the potential of deploying autonomous RISs in wireless networks with essentially no network overhead.
\end{abstract}


%
\IEEEpeerreviewmaketitle

\section{Introduction}
\label{sec_introduction}
A reconfigurable intelligent surface (RIS) is an innovative technology that has the ability to shape a wireless channel in beneficial ways thanks to the use of adjustable reflecting elements~\cite{Basar2019Wireless,Marco2020Smart}. They can be used for various purposes, such as improving network throughput, coverage, or energy efficiency. It is commonly assumed that RISs are essentially passive arrays without radio-frequency (RF) chains and computing resources, and are fully controlled by an external entity such as a base station (BS). To reap the benefits of RISs, channel state information (CSI) is also generally required. However, CSI estimation in passive RIS systems is challenging, often requiring a high pilot overhead that can significantly reduce the spectral efficiency~\cite{Cunhua2022Overview,Kim2022PracIRS}. In addition, the requirement of a control link through a wired cable or wireless channel limits the deployment flexibility of RISs and increases the complexity of the system installation, configuration, and maintenance costs. In some circumstances, establishing such a control link may be infeasible, for example, when an RIS is only temporarily deployed to an area.

\begin{figure}
    \centering
    \includegraphics[width = 1.0
    \linewidth]{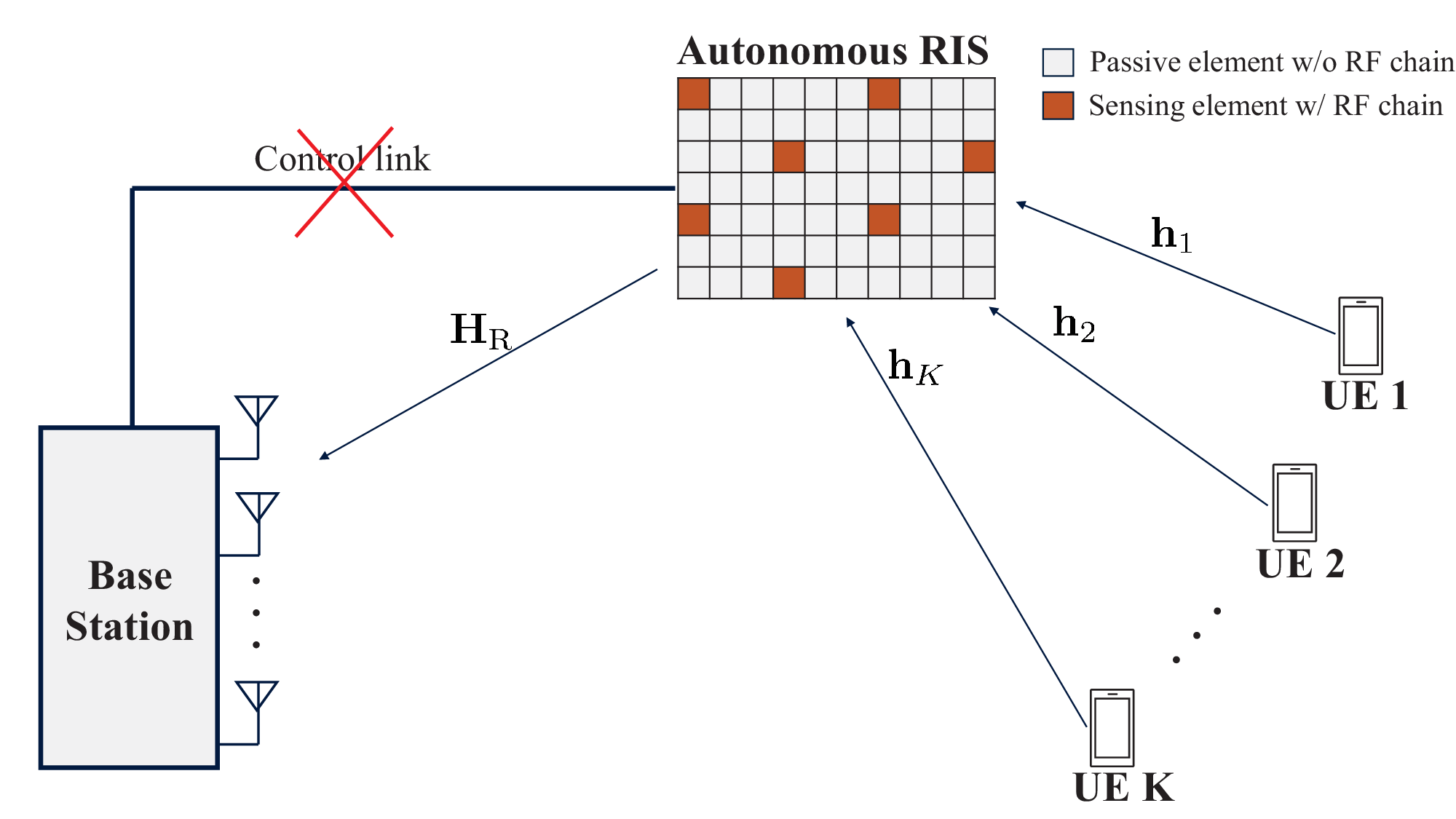}
    \caption{Illustration of an autonomous RIS-assisted system.}
    \label{fig:system-model}
\end{figure}

Recently, hybrid RIS structures have been introduced in which the RIS elements are able to simultaneously reflect and sense the incoming 
signal~\cite{Alexandropoulos2021Hybrid,Alamzadeh2021Reconfigurable,Zhang2022Channel}. Such structures pave the way for a new methodology where an RIS can ``sense-then-shape'' the environment itself. The benefit of hybrid RISs in terms of pilot overhead reduction has been recently reported in~\cite{Zhang2022Channel,Ly2023DD}. Motivated by these recent advances, we study the concept of an autonomous RIS as depicted in Fig.~\ref{fig:system-model}, which is \textit{self-configured} instead of being fully controlled by a remote BS, thus maximizing the deployment flexibility as well as simplifying the system configuration. Since an autonomous RIS is self-configured, the BS can operate using conventional protocols, e.g., estimating the effective instantaneous CSI and then performing combining/precoding. This approach is thus more practical than the one that the BS estimates the instantaneous cascaded CSI, jointly optimizes the combining/precoding and the phase shifts of the RIS, and then forwards the solution to the RIS via a control link \cite{taha2020,Wang2022IRS}.

The proposed autonomous RIS employs a hybrid RIS with some sensing elements and processing capability to self-configure its phase shifts using a deep Q network (DQN) \cite{mnih2013playing}. DQN is a reinforcement learning method that is useful when the state, e.g., a received signal, is strongly related to the channel environment, and the action, e.g., the RIS configuration, changes the channel environment. Since it is impractical to define the received signals and wireless channels in discrete sets, DQN is widely used for communication system developments, where a deep neural network is used to create a continuous-valued Q table \cite{taha2020,mismar2020,Wang2022IRS}. Without any control link, the partial observations provided by the RIS sensing elements are the only information available to the autonomous RIS. The key contribution of the paper is a method to convert the partial observations of the hybrid RIS into an estimate of the sum rate, which serves as the reward or Q value of the reinforcement learning-based DQN.

The paper is organized as follows. In Section~\ref{sec2}, we present a cluster-based channel model suitable for describing time variations due to changing positions of the clusters and mobile user equipment (UE). Section~\ref{sec3} provides the RIS system model and defines the observations from a few RIS sensing elements. Using the RIS observations, we develop a method for evaluating the sum-rate in Section~\ref{sec4}. Section~\ref{sec5} details the proposed DQN approach, and Section~\ref{sec_results} illustrates its performance using simulation results. Concluding remarks follow in Section~\ref{sec_conclusion}.

\textit{Notation:} Upper-case and lower-case boldface letters are used to indicate matrices and column vectors, respectively. The element-wise absolute value of vector $\mbf{a}$ is represented as $|\mbf{a}|$. The $p$-norm of vector $\mbf{a}$ is $\lVert\mbf{a}\rVert_p$. The indicator operator $[\![\mbf{a}]\!]_\infty$ is defined as $[\![\mbf{a}]\!]_p=[a_1^p\ a_2^p\ \cdots]^\mathrm{T}/\lVert\mbf{a}\rVert_p^p$ for $p\to\infty$. The $i$-th element of vector $\mbf{a}$ and the $i$-th row of matrix $\mbf{A}$ are given as $[\mbf{a}]_i$ and $[\mbf{A}]_{a,:}$, respectively. A partial vector formed from elements of vector $\mbf{a}$ using the set of indices $\mathcal{I}$ is defined as $[\mbf{a}]_\mathcal{I}$. The matrices $\mbf{A}^\mathrm{T}$ and $\mbf{A}^\mathrm{H}$ denote the transpose and the conjugate transpose of $\mbf{A}$, respectively. The discrete Fourier transform (DFT) of matrix $\mbf{A}$ is given by $\text{DFT}\{\mbf{A}\}$, where the DFT is performed column-wise. The set of complex numbers is denoted as $\mathbb{C}$.

\section{Channel Model}
\label{sec2}
\begin{figure}
    \centering
    \includegraphics[width = 0.65\linewidth]{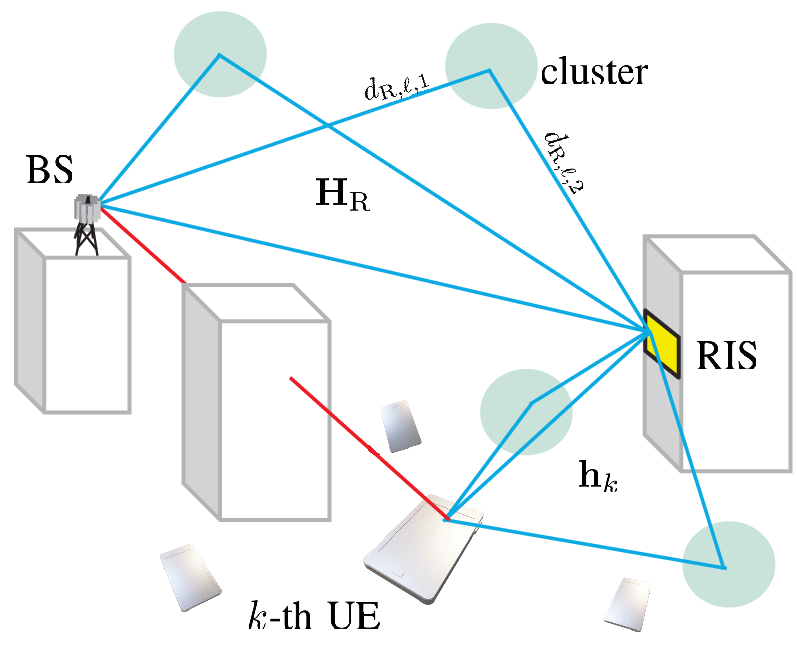}
    \caption{Assumed cluster-based channel model. The direct links between BS and UEs are blocked.}
    \label{fig:channel-model}
\end{figure}

We consider a system with an $M$-antenna BS serving $K$ single-antenna UEs with the help of an $N$-element RIS. We assume the cluster-based channel model depicted in Fig.~\ref{fig:channel-model}, where there is no direct channel between the UEs and the BS. Our model considers signals arriving at the autonomous RIS from both the BS and UEs. The channel between the $k$-th UE and the RIS is modeled as
\begin{align}\label{eq:ch_ue}
    \mbf{h}_k & = \sum_{\ell = 0}^{L_k-1} \frac{\sqrt{P_0}}{\prod_{p=0}^{P_{k,\ell}-1}(d_{k,\ell,p})^\alpha}e^{-j2\pi f_c\tau_{k,\ell}}\mbf{a}_{\RIS}(\theta_{k,\ell}^\text{hor},\theta_{k,\ell}^\text{ver}),
\end{align}
where $f_c$ is the carrier frequency, $\alpha$ is the pathloss exponent, $P_0$ is the reference channel power, and $L_k$ is the number of distinct channel paths, each of which can arrive via multiple reflections. The delay $\tau_{k,l}$ of channel path $\ell$ is given by $\tau_{k,\ell}=\sum_{p=0}^{P_{k,\ell}-1}d_{k,\ell,p}/c$ where $c$ is the speed of light, and $d_{k,\ell,p}$ represents the length of the $p$-th segment of the $\ell$-th channel path for the $k$-th UE. The uniform planar array (UPA) response at the RIS is defined as
\begin{align}\label{eq:arr_resp}
    & \mbf{a}_\RIS(\theta_{k,\ell}^\text{hor},\theta_{k,\ell}^\text{ver})= \nonumber \\
    & \qquad [1\ e^{j\pi\cos(\theta_{k,\ell}^\text{hor})\sin(\theta_{k,\ell}^\text{ver})}\ \cdots\ e^{j(N_\text{hor}-1)\pi\cos(\theta_{k,\ell}^\text{hor})\sin(\theta_{k,\ell}^\text{ver})}]^\mathrm{T} \nonumber \\
    & \qquad \otimes [1\ e^{j\pi\cos(\theta_{k,\ell}^\text{ver})}\ \cdots\ e^{j(N_\text{ver}-1)\pi\cos(\theta_{k,\ell}^\text{ver})}]^\mathrm{T},
\end{align}
where $\theta_{k,\ell}^\text{hor}$ and $\theta_{k,\ell}^\text{ver}$ are the zenith and azimuth angles of arrival (ZoA/AoA) for the $\ell$-th channel path of the $k$-th UE.

The BS-to-RIS channel is modeled as
\begin{align}\label{eq:ch_bs}
    \mbf{H}_\mathrm{R} & = \sum_{\ell=0}^{L_\mathrm{R}-1}\frac{\sqrt{P_0}}{\prod_{p=0}^{P_{\mathrm{R},\ell}-1}(d_{\mathrm{R},\ell,p})^\alpha}e^{-j2\pi f_c \tau_{\mathrm{R},\ell}} \nonumber \\
    & \qquad\qquad\qquad\qquad\times\mbf{a}_\mathrm{R}(\theta_{\mathrm{R},\ell}^\text{hor},\theta_{\mathrm{R},\ell}^\text{ver})\mbf{a}_\mathrm{B}^\mathrm{H}(\phi_{\mathrm{R},\ell}^\text{ver})
\end{align}
where the segment lengths $d_{\mathrm{R},\ell,p}$ and the delay $\tau_{\mathrm{R},\ell}$ are defined similarly to \eqref{eq:ch_ue}. We assume a uniform linear array (ULA) at the BS whose response $\mbf{a}_\mathrm{B}(\cdot)\in\mathbb{C}^{M\times 1}$ can be found as in \eqref{eq:arr_resp} with $N_\text{hor}=1$, and the ZoA/AoA and zenith of departure (ZoD) are $\theta_{\mathrm{R},\ell}^\text{hor},\theta_{\mathrm{R},\ell}^\text{ver}$ and $\phi_{\mathrm{R},\ell}^\text{ver}$, respectively. The positions of the clusters, BS, RIS, and UE nodes define the parameters of the channels, including the lengths of the channel segments and the ZoDs, AoAs, and ZoDs of the channel paths. Time-variations in the positions of the clusters and UEs produce channel variations. In Section~\ref{sec_results}, we describe the model assumed for cluster and UE motion in the numerical studies.

\section{System Model}
\label{sec3}
The effective uplink (UL) channel for the $k$-th UE is given as
\begin{equation}
    \boldsymbol{\mathfrak{h}}_{k} = \mbf{H}_\mathrm{R}^\mathrm{H}\diag{(\mbf{v})}\mbf{h}_k,
\end{equation}
where $\mbf{v}$ is the RIS phase shift vector defined as $\mbf{v}=[e^{j\psi_1}$ $\cdots\ e^{j\psi_N}]^\mathrm{T}$.
Let $\mbf{H} = [\boldsymbol{\mathfrak{h}}_1,\ldots,\boldsymbol{\mathfrak{h}}_K] \in \mbb{C}^{M \times K}$ be the effective uplink channel matrix. We assume reciprocity, so the downlink channel matrix is $\mbf{H}^\mathrm{H} = [\boldsymbol{\mathfrak{h}}_1,\ldots,\boldsymbol{\mathfrak{h}}_K]^\mathrm{H} \in \mbb{C}^{K\times M}$.

In a time division duplex (TDD) system, the downlink (DL) signal received at the $k$-th UE is written as
\begin{align}
    r_k=\boldsymbol{\mathfrak{h}}_k^\mathrm{H}\mbf{F}\mbf{s}+n_k,
\end{align}
where the DL precoding matrix $\mbf{F}$ consists of $K$ precoding vectors $\mbf{F}=[\mbf{f}_1\ \cdots\ \mbf{f}_K]$, and the DL symbol vector is given as $\mbf{s}=[s_1\ \cdots\ s_K]^\mathrm{T}$. The noise at the $k$-th UE is zero-mean Gaussian with variance $\sigma_n^2$, which we denote as $n_k\sim\mathcal{CN}(0,\sigma_n^2)$. We assume that the DL symbols are randomly generated and satisfy $\mathbb{E}[\mbf{s}\mbf{s}^\mathrm{H}]=P_\text{BS}\mbf{I}_K$, where $P_\text{BS}$ is the BS transmit power, and $\mbf{I}_K$ is a $K\times K$ identity matrix.

 The limited number of RIS sensing elements enables the RIS to obtain partial information about the UL and DL channels, The signal from the BS to the RIS is given
 as
\begin{equation}\label{eq:rx_bs_sp1}
    [\mbf{y}_\mathrm{R}]_{\mathcal{I}_s}=[\mbf{H}_\mathrm{R}]_{\mathcal{I}_s,:}\mbf{F}\mbf{s}+\mbf{n}_{\mathrm{R},\mathcal{I}_s},
\end{equation}
where $\mathcal{I}_s$ is the set of indices corresponding to the sensing elements, the noise term $\mbf{n}_{\mathrm{R},\mathcal{I}_s}$ is distributed as $\mathcal{CN}(\boldsymbol{0},\sigma_n^2\mbf{I}_{|\mathcal{I}_s|})$, and $|\mathcal{I}_s|$ is the cardinality of $\mathcal{I}_s$. The received signal at the RIS from the $k$-th UE is given by
\begin{equation}
    [\mbf{y}_k]_{\mathcal{I}_s}=[\mbf{h}_k]_{\mathcal{I}_s}x_k+\mbf{n}_{k,\mathcal{I}_s},
\end{equation}
where $x_k$ is the UL symbol from the $k$-th UE with transmit power $\mathbb{E}[|x_k|^2]=P_\text{UE}$. The noise $\mbf{n}_{k,\mathcal{I}_s}$ is also assumed to be distributed as $\mathcal{CN}(\mbf{0},\sigma_n^2\mbf{I}_{|\mathcal{I}_s|})$.

In the following section, we propose a method for evaluating the sum-rate using only the partial observations from the RIS sensing elements. The derived sum-rate will then be used to form the DQN reward.

\section{Proposed Sum-Rate Evaluation}
\label{sec4}
\subsection{Observation Recovery}\label{sec4_1}
The full-dimensional RIS received signals $\mbf{y}_\mathrm{R}$ from the BS and $\mbf{y}_k$ from the $k$-th UE are necessary for the sum-rate evaluation. From \eqref{eq:ch_bs}, $\mbf{y}_\mathrm{R}$ can be represented as
\begin{align}\label{eq:rx_bs}
    \mbf{y}_\mathrm{R} & =\mbf{H}_\mathrm{R}\mbf{F}\mbf{s}+\mbf{n}_\mathrm{R}\nonumber \\
    & =\sum_{\ell=0}^{L_\mathrm{R}-1}\frac{\sqrt{P_0}}{\prod_{p=0}^{P_{\mathrm{R},\ell}-1}(d_{\mathrm{R},\ell,p})^\alpha}e^{-j2\pi f_c\tau_{\mathrm{R},\ell}} \nonumber \\
    &\qquad\qquad\qquad\qquad\times\mbf{a}_\mathrm{R}(\theta_{\mathrm{R},\ell}^\text{hor},\theta_{\mathrm{R},\ell}^\text{ver})\mbf{a}_\mathrm{B}^\mathrm{H}(\phi_{\mathrm{R},\ell}^\text{ver})\mbf{F}\mbf{s}+\mbf{n}_\mathrm{R} \nonumber \\
    & = \sum_{\ell=0}^{L_\mathrm{R}-1}\beta_{\mathrm{R},\ell}\mbf{a}_\mathrm{R}(\theta_{\mathrm{R},\ell}^\text{hor},\theta_{\mathrm{R},\ell}^\text{ver})+\mbf{n}_\mathrm{R},
\end{align}
by introducing $\beta_{\mathrm{R},\ell}=\frac{\sqrt{P_0}}{\prod_{p=0}^{P_{\mathrm{R},\ell}-1}(d_{\mathrm{R},\ell,p})^\alpha}e^{-j2\pi f_c\tau_{\mathrm{R},\ell}}\mbf{a}_\mathrm{B}^\mathrm{H}(\phi_{\mathrm{R},\ell}^\text{ver})\mbf{F}\mbf{s}$. In \eqref{eq:rx_bs}, the parameters $\beta_{\mathrm{R},\ell},\theta_{\mathrm{R},\ell}^\text{hor}$, and $\theta_{\mathrm{R},\ell}^\text{ver}$ fully define the channel. The partial observation in \eqref{eq:rx_bs_sp1} can be written as
\begin{align}\label{eq:rx_bs_sp2}
    [\mbf{y}_\mathrm{R}]_{\mathcal{I}_s}=\sum_{\ell=0}^{L_\mathrm{R}-1}\beta_{\mathrm{R},\ell}[\mbf{a}_\mathrm{R}(\theta_{\mathrm{R},\ell}^\text{hor},\theta_{\mathrm{R},\ell}^\text{ver})]_{\mathcal{I}_s}+\mbf{n}_{\mathrm{R},\mathcal{I}_s},
\end{align}
which still contains information about $\beta_{\mathrm{R},\ell},\theta_{\mathrm{R},\ell}^\text{hor}$, and $\theta_{\mathrm{R},\ell}^\text{ver}$. If the channel parameters $\beta_{\mathrm{R},\ell},\theta_{\mathrm{R},\ell}^\text{hor}$, and $\theta_{\mathrm{R},\ell}^\text{ver}$ can be extracted from \eqref{eq:rx_bs_sp2}, it is possible to reconstruct $\eqref{eq:rx_bs}$ assuming the noise is not excessive.

The ZoA/AoA $\theta_{\mathrm{R},\ell}^\text{hor}$ and $\theta_{\mathrm{R},\ell}^\text{ver}$ can be inferred using the orthogonal matching pursuit (OMP) algorithm \cite{OMP}. The autocorrelation matrix for OMP can be expressed as
\begin{equation}
    \mbf{R}_{\mbf{y}_\mathrm{R}}=\mathbb{E}\left[[\mbf{y}_\mathrm{R}]_{\mathcal{I}_s}[\mbf{y}_\mathrm{R}]_{\mathcal{I}_s}^\mathrm{H}\right],
\end{equation}
which can be obtained by a sample average. The OMP algorithm gives the estimated ZoA/AoA $\hat{\theta}_{\mathrm{R},\ell}^\text{hor}$ and $\hat{\theta}_{\mathrm{R},\ell}^\text{ver}$ from $\mbf{R}_{\mbf{y}_\mathrm{R}}$. With AoAs obtained by OMP, the remaining channel parameter $\beta_{\mathrm{R},\ell}$ can be computed as
\begin{align}
    \hat{\beta}_{\mathrm{R},\ell} & =\frac{1}{|\mathcal{I}_s|}[\mbf{a}(\hat{\theta}_{\mathrm{R},\ell}^\text{hor},\hat{\theta}_{\mathrm{R},\ell}^\text{ver})]_{\mathcal{I}_s}^\mathrm{H}[\mbf{y}_\mathrm{R}]_{\mathcal{I}_s} \nonumber \\
    & \approx \beta_{\mathrm{R},\ell}+\frac{1}{|\mathcal{I}_s|}[\mbf{a}(\hat{\theta}_{\mathrm{R},\ell}^\text{hor},\hat{\theta}_{\mathrm{R},\ell}^\text{ver})]_{\mathcal{I}_s}^\mathrm{H}\mbf{n}_{\mathrm{R},\mathcal{I}_s}.
\end{align}
With the estimated parameters $\hat{\theta}_{\mathrm{R},\ell}^\text{hor},\hat{\theta}_{\mathrm{R},\ell}^\text{ver}$ and $\hat{\beta}_{\mathrm{R},\ell}$, the full-dimensional RIS received signal from the BS is recovered as
\begin{equation}
    \hat{\mbf{y}}_\mathrm{R} = \sum_{\ell=0}^{L_\mathrm{R}-1}\hat{\beta}_{\mathrm{R},\ell}\mbf{a}_\mathrm{R}(\hat{\theta}_{\mathrm{R},\ell}^\text{hor},\hat{\theta}_{\mathrm{R},\ell}^\text{ver}).
\end{equation}
The full-dimensional received signal from the $k$-th UE can be obtained using the same procedure.

\begin{figure*}
    \centering
    \includegraphics[width=0.73\linewidth]{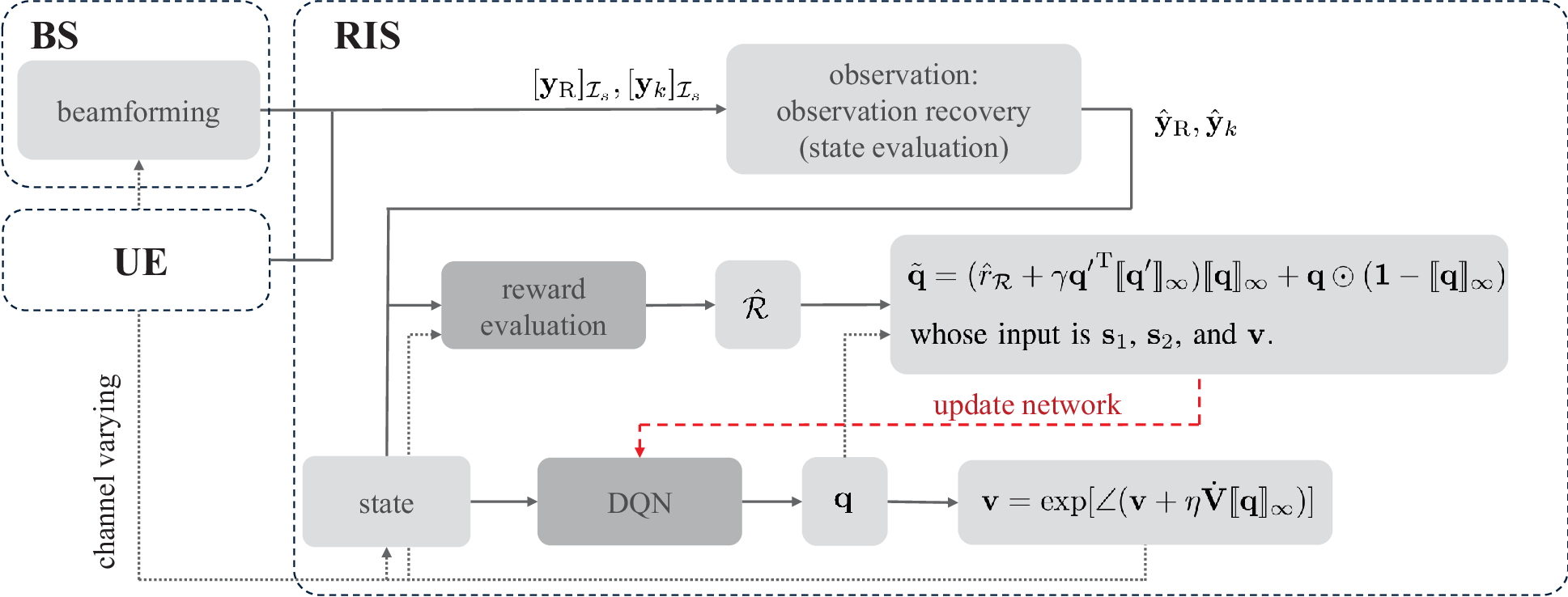}
    \caption{Flowchart of the proposed DQN interacting with the environment.}
    \label{fig:enter-label}
\end{figure*}

\subsection{Sum-Rate Evaluation Method}
The DL sum-rate defined as
\begin{align}\label{eq:sum_rate}
    \mathcal{R} =\sum_{k=1}^K\log_2\left(1+\frac{P_\text{BS}|\mathfrak{h}_k^\mathrm{H}\boldsymbol{\mbf{f}}_k|^2}{\sum_{k'\neq k}P_\text{BS}|\mathfrak{h}_k^\mathrm{H}\boldsymbol{\mbf{f}}_{k'}|^2+\sigma_n^2}\right)
\end{align}
can be used as the DQN reward. However, the autonomous RIS cannot directly calculate the sum-rate in \eqref{eq:sum_rate} since it requires precise CSI. The autonomous RIS therefore has to evaluate the sum-rate relying on the sensed observations from the RIS. 

We first define the observation
\begin{equation}
    z_k = \hat{\mbf{y}}_\mathrm{R}^\mathrm{H}\diag\{\mbf{v}\}\hat{\mbf{y}}_k
\end{equation}
with the RIS received signals $\hat{\mbf{y}}_\mathrm{R}$ and $\hat{\mbf{y}}_k$ that are recovered in Section \ref{sec4_1}. Assuming perfect recovery of $\mbf{y}_\mathrm{R}$ and $\mbf{y}_k$, the observation is expanded as
\begin{align}\label{eq:obsv}
    z_k = & (\mbf{H}_\mathrm{R}\mbf{F}\mbf{s}+\mbf{n}_\mathrm{R})^\mathrm{H}\diag\{\mbf{v}\}(\mbf{h}_k x_k + \mbf{n}_k) \nonumber \\
    = & \mbf{s}^\mathrm{H}\mbf{F}^\mathrm{H}\mbf{H}_\mathrm{R}^\mathrm{H}\diag\{\mbf{v}\}\mbf{h}_k x_k
    + \mbf{s}^\mathrm{H}\mbf{F}^\mathrm{H}\mbf{H}_\mathrm{R}^\mathrm{H}\diag\{\mbf{v}\}\mbf{n}_k \nonumber \\
    & + \mbf{n}_\mathrm{R}^\mathrm{H}\diag\{\mbf{v}\}\mbf{h}_k x_k
    +\mbf{n}_\mathrm{R}^\mathrm{H}\diag\{\mbf{v}\}\mbf{n}_k.
\end{align}
If we assume that the BS precoder is designed sufficiently well, e.g., using zero-forcing (ZF) or minimum mean square error (MMSE) precoders, the inter-user interference (IUI) can be assumed to be negligible, i.e., $\sum_{k'\neq k}\mbf{f}_{k'}^\mathrm{H}\mbf{H}_\mathrm{R}^\mathrm{H}\diag\{\mbf{v}\}\mbf{h}_k\approx 0$. With this assumption, \eqref{eq:obsv} can be further formulated as
\begin{align}
    z_k \approx & s_k^\ast\mbf{f}_k^\mathrm{H}\mbf{H}_\mathrm{R}^\mathrm{H}\diag\{\mbf{v}\}\mbf{h}_k x_k
    + s_k^\ast\mbf{F}^\mathrm{H}\mbf{H}_\mathrm{R}^\mathrm{H}\diag\{\mbf{v}\}\mbf{n}_k \nonumber \\
    & + \mbf{n}_\mathrm{R}^\mathrm{H}\diag\{\mbf{v}\}\mbf{h}_k x_k
    +\mbf{n}_\mathrm{R}^\mathrm{H}\diag\{\mbf{v}\}\mbf{n}_k,
\end{align}
from which the following can be evaluated:
\begin{align}\label{eq:expect_z}
    \mathbb{E}[|z_k|^2] & \approx P_\text{BS}P_\text{UE}|\mbf{f}_k^\mathrm{H}\mbf{H}_\mathrm{R}^\mathrm{H}\diag\{\mbf{v}\}\mbf{h}_k|^2+N\sigma_n^4 \nonumber \\
    & = P_\text{BS}P_\text{UE}|\mbf{f}_k^\mathrm{H}\boldsymbol{\mathfrak{h}}_k|^2 + N\sigma_n^4,
\end{align}
where the expectation is taken over the DL symbol $s_k$, UL symbol $x_k$, and noise signals $\mbf{n}_\mathrm{R}$ and $\mbf{n}_k$ in the DL and UL transmissions, respectively. The expectation in \eqref{eq:expect_z} can be evaluated using a sample average. Finally, the sum-rate can be evaluated as
\begin{align}\label{eq:sum_rate_ev}
    \hat{\mathcal{R}} & =\sum_{k=1}^K\log_2\left(1+\frac{\mathbb{E}[|z_k|^2]/P_\text{UE}}{\sigma_n^2}\right) \nonumber \\
    & = \sum_{k=1}^K\log_2\left(1+\frac{P_\text{BS}|\boldsymbol{\mathfrak{h}}_k^\mathrm{H}\mbf{f}_k|^2+N\sigma_n^4/P_\text{UE}}{\sigma_n^2}\right).
\end{align}
The sum-rate in \eqref{eq:sum_rate_ev} will be close to the actual sum-rate in \eqref{eq:sum_rate} assuming a proper beamformer $\mbf{F}$ satisfying $\sum_{k'\neq k}|\boldsymbol{\mathfrak{h}}_k^\mathrm{H}\mbf{f}_{k'}|^2 \approx 0$ and sufficient transmit power $P_\text{BS}\gg \sigma_n^2$. Section~\ref{sec5} investigates more details of the proposed DQN, where the sum-rate evaluated in \eqref{eq:sum_rate_ev} is used to define a reward.

\section{Proposed DQN Design} \label{sec5}
DQN is a reinforcement learning method with a deep neural network. A flowchart for our proposed DQN is given in Fig.~\ref{fig:enter-label}. The RIS recovers the received signal from its sensing elements and evaluates the sum-rate. The recovered observation is transformed into the DQN state, the input of the DQN neural network, and the sum-rate gives the target Q value. The DQN neural network outputs the Q value, after which the target Q value is updated and accumulated for DQN training. The RIS chooses the action according to a given policy, and then the RIS phase shifts are updated. As a result, the RIS channel changes, and the BS modifies the beamformer for the given RIS channel. We present further details about the arguments and structure of the DQN in Sections~\ref{sec5_1} and~\ref{sec5_2}, respectively. The DQN training process is then explained in Section~\ref{sec5_3}.

\subsection{DQN Arguments}\label{sec5_1}
\subsubsection{Environment} The environment refers to the medium with which the agent interacts, including the BS, the wireless channels, and the UEs.
\subsubsection{Agent} The RIS acts as the agent.
\subsubsection{State} The DQN has two states including the combined RIS observation
\begin{align}\label{eq:state1}
    \mbf{s}_{1,k}
    = \left\lvert\text{DFT}\left\{\diag\{\mbf{y}_\mathrm{R}^\mathrm{H}\}\diag\{\mbf{v}\}\mbf{y}_k\right\}\right\rvert^\mathrm{T},
\end{align}
from which we obtain $\mbf{s}_1=[\mbf{s}_{1,1}\ \cdots\ \mbf{s}_{1,K}]$, and the RIS phase shift vector
\begin{align}\label{eq:state2}
    \mbf{s}_2 = \left\lvert\text{DFT}\left\{\mbf{v}\right\}\right\rvert^\mathrm{T},
\end{align}
which are the DQN inputs that go into separate pipelines of the DQN structure and merge to predict the Q value. The combined observation in \eqref{eq:state1} can be represented as
\begin{align}\label{eq:state1_approx}
    \mbf{s}_{1,k}
    &\approx \left\lvert\text{DFT}\left\{\diag\{s_k^\ast\mbf{f}_k^\mathrm{H}\mbf{H}_\mathrm{R}^\mathrm{H}\}\diag\{\mbf{v}\}\mbf{h}_kx_k\right\}\right\rvert^\mathrm{T} \nonumber \\
    &= \left\lvert s_k^\ast x_k\text{DFT}\left\{\diag\{\mbf{f}_k^\mathrm{H}\mbf{H}_\mathrm{R}^\mathrm{H}\}\diag\{\mbf{v}\}\mbf{h}_k\right\}\right\rvert^\mathrm{T} \nonumber \\
    &=\left\lvert s_k^\ast x_k \text{DFT}\left\{\diag\{\mbf{f}_k^\mathrm{H}\}\boldsymbol{\mathfrak{h}}_k\right\}\right\rvert^\mathrm{T} \nonumber \\
    &=\left\lvert s_k^\ast x_k \diag\{\mbf{f}_k^\mathrm{H}\}\text{DFT}\left\{\boldsymbol{\mathfrak{h}}_k\right\}\right\rvert^\mathrm{T}
\end{align}
where the approximation is due to the assumptions $\sum_{k'\neq k}|\boldsymbol{\mathfrak{h}}_k^\mathrm{H}\mbf{f}_{k'}|^2 \approx 0$, and $P_\text{BS}\gg \sigma_n^2$. Since the representation in \eqref{eq:state1_approx} becomes a function of the concatenated RIS channel $\boldsymbol{\mathfrak{h}}_k$, we use the combined observation as the state. Note that the DFT operations in \eqref{eq:state1} and \eqref{eq:state2} convert the combined RIS observation $\diag\{\mbf{y}_\mathrm{R}^\mathrm{H}\}\diag\{\mbf{v}\}\mbf{y}_k$ and the RIS phase shift vector $\mbf{v}$ into the spatial domain. This transformation makes the signal sparser and more informative to the DQN. Since the absolute value varies slowly in the spatial domain and is robust to overfitting, we take the absolute values for each state.

\subsubsection{Action} The action chosen by the policy determines the RIS phase update. The optimal policy selects the maximum Q value as follows:
\begin{equation}\label{eq:policy}
    a^\ast = \argmax_{a\in\mathcal{A}} Q^\ast([\mbf{s}_1,\mbf{s}_2],a),
\end{equation}
where $\mathcal{A}=\{1,\dots,N_a\}$ represents the set of $N_a$ possible actions, and $Q^\ast([\mbf{s}_1,\mbf{s}_2],a)$ is the action-value function. We define a vector $\mbf{q}$ whose $a$-th element is $Q^\ast([\mbf{s}_1,\mbf{s}_2],a)$, and with the states $[\mbf{s}_1,\mbf{s}_2]$ and the Q value vector $\mbf{q}$, we can define the DQN as
\begin{equation}\label{eq:Qfunc}
    \mbf{q} = f_\text{DQN}(\mbf{s}_1,\mbf{s}_2).
\end{equation}
The RIS phase shifts are updated as
\begin{equation}\label{eq:phs_update}
    \mbf{v}=\exp[j\angle(\mbf{v}+\eta\Delta\mbf{v})],
\end{equation}
where the update direction is defined as $\Delta\mbf{v}=\dot{\mbf{V}}\mbf{e}_{a^\ast}$, $\eta$ is the step size, the action set matrix $\dot{\mbf{V}}=[\dot{\mbf{v}}_1\ \cdots\ \dot{\mbf{v}}_{N_a}]$ is a set of possible update directions for the RIS phase shifts, and $\mbf{e}_{a^\ast}$ is a one-hot vector whose only non-zero element is in position $a^\ast$ and is equal to 1. In particular, we set $\mbf{e}_{a^\ast}=[\![\mbf{q}]\!]_\infty$. In our implementation, we will choose the action set matrix to be a DFT matrix. Since the columns of the DFT matrix form the basis of an $N$-dimensional space, consecutive action decisions can be interpreted as different linear combinations of the basis vectors. Thus, even with a discrete action set, the resulting RIS phase shift can be any vector in $N$-dimensional space. To tune the convergence of the RIS update, we use the following adaptive learning rate 
\begin{align}
    \eta=0.1\lvert\Delta\mathcal{R}\rvert+0.01,
\end{align}
where $\Delta\mathcal{R}=\mathcal{R}'-\mathcal{R}$ is the rate increment, and serves as a proxy for the gradient.

\begin{figure}
    \centering
    \includegraphics[width=0.9\linewidth]{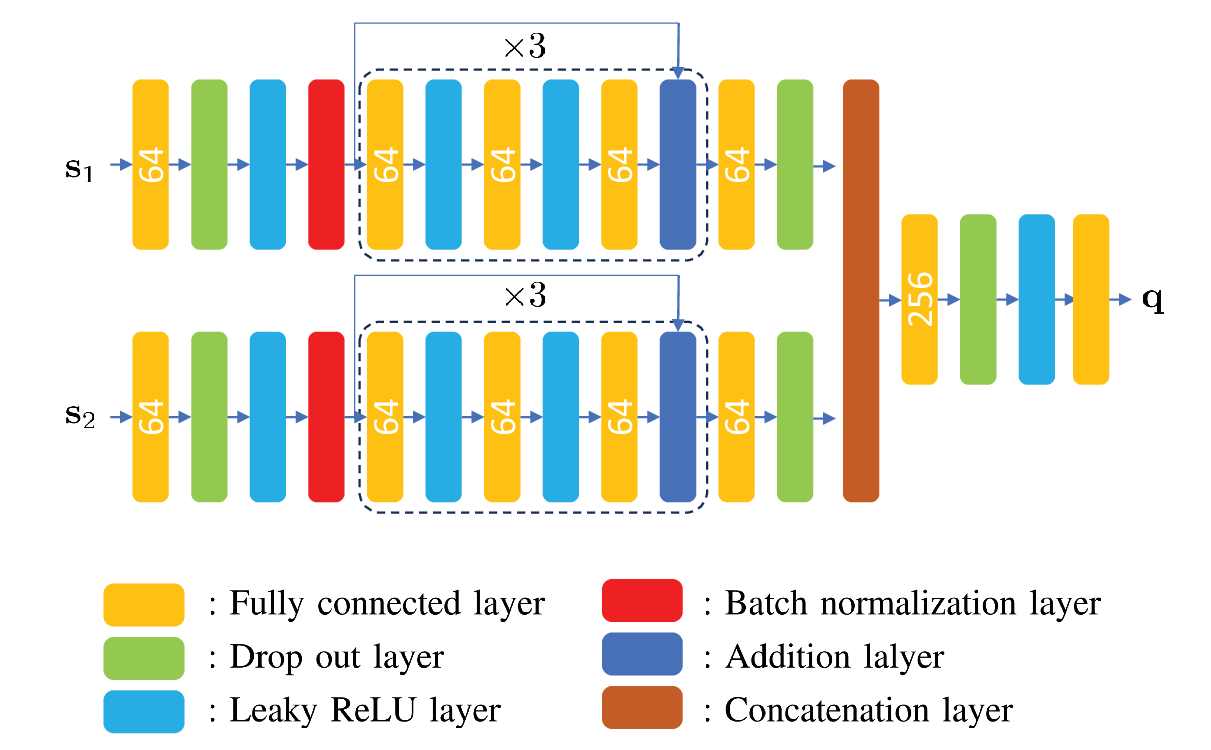}
    \caption{Structure of the DQN neural network with two input pipelines. The number of weight parameters is written on each layer.}
    \label{fig:DQNstruct}
\end{figure}

\subsubsection{Policy} We employ an $\epsilon$-greedy policy. During the initial stages, the Q value is not 
reliable, so choosing the best action as in \eqref{eq:policy} is not the best strategy. Thus initially, with probability $\epsilon$, the action is randomly chosen. The value of $\epsilon$ gradually decreases so that the best action in \eqref{eq:policy} is more likely to be selected in later stages.

\subsubsection{Reward} Instead of directly using the sum-rate as the reward, we propose to define the reward as the sum-rate ratio $\hat{r}_\mathcal{R}=\hat{\mathcal{R}}'/\hat{\mathcal{R}}$, where $\hat{\mathcal{R}}$ is the present estimated sum-rate and $\hat{\mathcal{R}}'$ is the next estimated sum-rate. Maximizing the sum-rate ratio is equivalent to maximizing the sum-rate, but it is more robust to arbitrary scaling.

The DQN neural network works as a Q table for conventional reinforcement learning. Since there are no labeled data for the reinforcement learning, the DQN training needs the target Q data. With the reward, the target Q value vector $\tilde{\mbf{q}}$ can be evaluated as
\begin{align}\label{eq:Qupdate}
    [\tilde{\mbf{q}}]_{a^\ast} = \hat{r}_\mathcal{R}+ \gamma \max_{a'} Q^\ast([\mbf{s}_1',\mbf{s}_2'],a')
    = \hat{r}_\mathcal{R}+ \gamma {\mbf{q}'}^\mathrm{T}[\![\mbf{q}']\!]_\infty,
\end{align}
where $\gamma$ is the discount factor, and $\mbf{q}'$ is the Q value vector for the next state $[\mbf{s}_1',\mbf{s}_2']$. The other elements in $\tilde{\mbf{q}}$ are equal to those in $\mbf{q}$. The target Q value vector $\tilde{\mbf{q}}$ is the desired DQN output for the next state, which trains the DQN $f_\text{DQN}(\cdot)$. However, if the $a$-th elements of $\tilde{\mbf{q}}$ and $f_\text{DQN}(\mbf{s}_1,\mbf{s}_2)$ are not of the same scale, which means $[\tilde{\mbf{q}}]_a$ is too large or too small, the target Q value vector $\tilde{\mbf{q}}$ might not be useful for training. Thus, we propose to use the following normalized target Q vector:
\begin{align}
    \check{\mbf{q}} = \frac{\eta}{\sigma_{\tilde{\mbf{q}}}}(\tilde{\mbf{q}}-m_{\tilde{\mbf{q}}}\mbf{1})+\frac{1}{1-\gamma},
\end{align}
where $m_{\tilde{\mbf{q}}}$ and $\sigma_{\tilde{\mbf{q}}}^2$ are the sample mean and sample variance of the elements in $\tilde{\mbf{q}}$. We still use the learning rate $\eta$ here to change the variance such that $\check{\mbf{q}}$ becomes comparable to $f_\text{DQN}(\mbf{s}_1,\mbf{s}_2)$. The loss function for training DQN is defined as
\begin{align} \text{Loss}=\lvert[\check{\mbf{q}}-f_\text{DQN}(\mbf{s}_1,\mbf{s}_2)]_{a^\ast}\rvert^2,
\end{align}
where only the difference for the $a^\ast$-th element gives the DQN loss for the backpropagation.

\subsection{DQN structure} \label{sec5_2}
The autonomous RIS exploits the DQN structure in Fig.~\ref{fig:DQNstruct} which is similar to \cite{Wang2022IRS}, but the component layers are slightly different. The DQN input states are the RIS observations and the RIS phase shift vector in Eqs. \eqref{eq:state1} and \eqref{eq:state2}, and the output is the Q value vector $\mbf{q}$ whose label can be obtained as in~\eqref{eq:Qupdate}. Since gradient descent is a critical issue for the DQN training, we add a batch normalization layer. The dropout layers are connected to the fully connected layers to address the overfitting issue. The activation layer here uses the leaky rectified linear unit (ReLU) function since the rate decrement should be considered in the DQN training and propagated in a negative direction \cite{xu2015empirical}.

\begin{figure}
    \centering
    \includegraphics[width=0.8\linewidth]{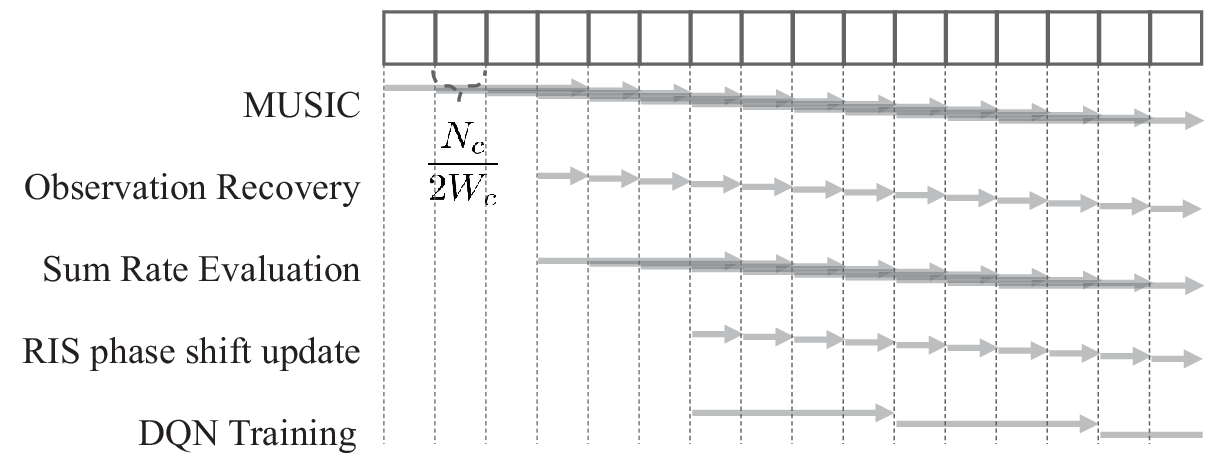}
    \caption{The procedure associated with the proposed DQN for autonomous RIS. The training sample period is $N_c/2W_s$, where $W_s$ is baseband bandwidth, and $N_c$ is the sampling interval.}
    \label{fig:proc_seq}
\end{figure}

\subsection{DQN Training} \label{sec5_3}
The proposed DQN presented above is an online machine learning approach, i.e., the DQN keeps updating its parameters using consecutive observations obtained from the RIS sensing elements. The processing sequence is shown in Fig.~\ref{fig:proc_seq}. The successively accumulated data provides the expected value of the data required for every time slot. The observation recovery and the RIS phase shift update employ instantaneous data from every training sample period. As a result, the DQN state and reward can be continuously evaluated to determine the necessary action. The DQN training is, however, not conducted with every sample. The training data is stacked for a given period and shuffled for the DQN training, which prevents overfitting. The computational complexity of the DQN is largely influenced by the training process. However, since actions can be selected concurrently during training, the proposed system is relatively unburdened by computational delay.

\section{Simulation Results}
\label{sec_results}

In this section, we present simulation results to show the effectiveness of our proposed DQN-based approach for the considered autonomous RIS system. The BS and the RIS are respectively located at $(0,0,35)$ m and $(-50,0,10)$ m, and the UEs are randomly placed in an area of size $100\times 50$ m$^2$ centered at $(0,0)$ m. There are two UEs, and the UE height is fixed to 1 m. The size of the BS ULA is $4\times1$, and the size of the RIS UPA is $4\times 8$. The direct path between the BS and UEs is assumed to be blocked. The BS employs the MMSE precoder \cite{tse2005fundamentals}. The first row and column of the RIS UPA are assumed to be the sensing elements, for a total of 13 RIS receivers. The system bandwidth is 20 MHz, the BS transmit power is 30 dBm, and the UE transmit power is 10 dBm.

The BS-to-RIS and UE-to-RIS channels consist of both line-of-sight (LoS) and non-LoS (NLoS) channels, where the NLoS channels are generated using clusters with random scattering in an area of size $200\times100\times50$ m$^3$, as illustrated in Fig.~\ref{fig:channel-model}. There are three clusters for each BS-to-RIS and UE-to-RIS channel. The clusters and UEs are also assumed to move at a fixed speed along a linear trajectory in random directions.

As explained in Section~\ref{sec5}, the autonomous RIS needs the full-dimensional observations and the sum-rate to obtain the DQN state and reward. To investigate the importance of each stage, we compare three DQNs: \textit{i)} aRIS\_ref1, with noise-free observations and precisely known sum-rate, \textit{ii)} aRIS\_ref2, with noise-free observations and the estimated sum-rate in \eqref{eq:sum_rate_ev}, and \textit{iii)} aRIS, with the proposed evaluated sum-rate after observation recovery as explained in Section~\ref{sec4_1}. We also compare these approaches with the use of random RIS phase shifts, which does not require any channel information.

\begin{figure}
    \centering
    \includegraphics[width=\linewidth]{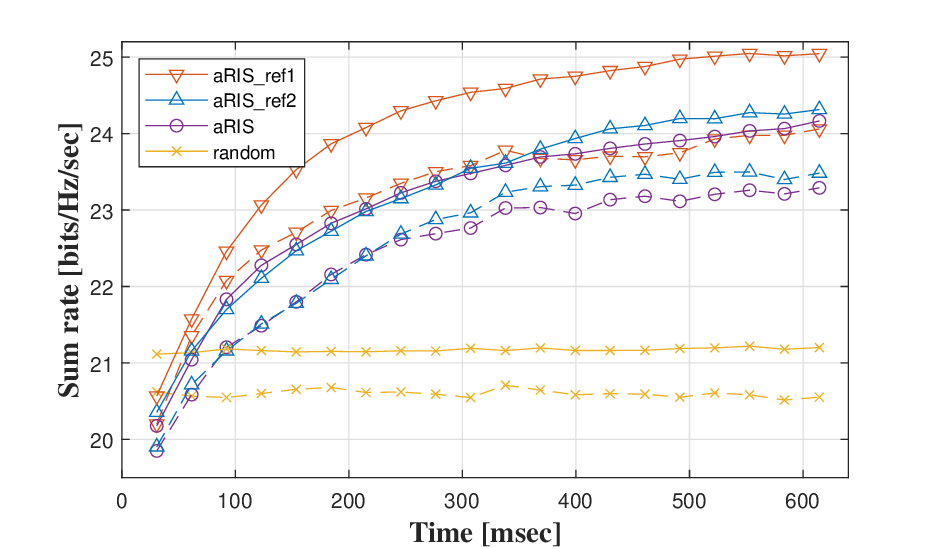}
    \caption{Sum-rates for DQNs with different levels of information. The bold and dotted lines are for the cases of 1 m/sec and 5 m/sec UE and cluster movement, respectively.}
    \label{fig:exp1}
\end{figure}

Fig.~\ref{fig:exp1} shows a sum rate comparison for two scenarios where the UEs and clusters move at speeds of 1 m/sec (bold lines) and 5 m/sec (dotted lines), respectively. The DQN for the case involving UE and cluster motion at 1 m/sec converges to a higher sum rate more rapidly than the case of 5 m/sec, since more rapidly varying channels are more challenging for DQN adaptation. For a given UE speed, aRIS\_ref1 achieves the best performance since it uses perfect knowledge of the channels, while the proposed a\_RIS is comparable to aRIS\_ref2 for both low and high mobility cases. This clearly shows the potential of autonomous RIS in practice.

\section{Conclusion}
\label{sec_conclusion}
In this paper, we have proposed an autonomous RIS that does not require external control. The autonomous RIS is equipped with a few sensing elements whose measured data are used by a DQN to find a self-configured RIS phase shift solution. The proposed DQN updates the RIS phase shifts in a way that enhances the sum-rate. Due to the relatively small number of RIS sensing elements, the DQN state and reward must be established using only partial observations. Simulation results show that even with limited sensing information, the proposed DQN can still enhance the RIS channel and outperform the a random RIS configuration. Although the proposed RIS system may not be as energy-efficient as RIS systems devoid of computational resources, leveraging task-oriented hardware such as neural network processing units (NPUs) could substantially reduce power consumption.

\section*{Acknowledgement}
This work was supported in part by the Korean Ministry of Science and ICT (MSIT) under the Information Technology Research Center support program (IITP-2020-0-01787) supervised by the Institute of Information \& Communications Technology Planning \& Evaluation, in part by the {U.S.} National Science Foundation under grants CNS-2107182 and ECCS-2030029, and in part by a Korea Institute for Advancement of Technology (KIAT) grant funded by the Ministry of Trade, Industry and Energy (MOTIE) through the International Cooperative R\&D program (P0022557).

\ifCLASSOPTIONcaptionsoff
  \newpage
\fi

\bibliographystyle{IEEEtran}
\bibliography{ref}


%









\end{document}